\documentclass[12pt]{iopart}

\usepackage{graphicx}
\usepackage[dvipdfmx,colorlinks=true,linkcolor=blue,citecolor=blue,urlcolor=blue]{hyperref}

\newcommand{\singlet}{{}^1\mathrm{S}_0}
\newcommand{\triplet}{{}^3\mathrm{P}_2}
\newcommand{\Li}{{}^2\mathrm{S}_{1/2}}

\newcommand{\cm}{\mathrm{cm}}
\newcommand{\micron}{\mu\mathrm{m}}
\newcommand{\nm}{\mathrm{nm}}
\newcommand{\msec}{\mathrm{ms}}
\newcommand{\s}{\mathrm{s}}
\newcommand{\G}{\mathrm{G}}
\newcommand{\mG}{\mathrm{mG}}
\newcommand{\eqref}[1]{equation~(\ref{#1})}

\begin{document}

\title[Spin dependent inelastic collisions]{Spin dependent inelastic
collisions between metastable state two-electron atoms and ground state
alkali-atoms}

\author{Florian Sch\"{a}fer$^1$, Hideki Konishi$^1$, Adrien Bouscal$^{1,2}$, Tomoya Yagami$^1$ and Yoshiro Takahashi$^1$}
\address{$^1$ Department of Physics, Graduate School of Science, Kyoto University, Kyoto 606-8502, Japan}
\address{$^2$ Department of Physics, \'{E}cole Normale Sup\'{e}rieure, 24 rue Lhomond, 75231 Paris Cedex 05, France}
\ead{schaefer@scphys.kyoto-u.ac.jp}
\date{\today}

\begin{abstract}
	Experimentally the spin dependence of inelastic collisions between ytterbium
	(Yb) in the metastable $\triplet$ state and lithium (Li) in the $\Li$ ground
	state manifold is investigated at low magnetic fields. Using selective
	excitation all magnetic sublevels $m_J$ of ${}^{174}$Yb($\triplet$) are
	accessed and four of the six lowest lying magnetic sublevels of ${}^6$Li are
	prepared by optical pumping. On the one hand, $m_J$-independence of
	collisions involving Li($F=1/2$) atoms is found. A systematic
	$m_J$-dependence in collisions with Li($F=3/2$) atoms, in particular
	suppressed losses for stretched collisional states, is observed on the other
	hand. Further, $m_J$-changing processes are found to be of minor relevance.
	The span of observed inelastic collision rates is between $1 \times
	10^{-11}$ and $40 \times 10^{-11}~\cm^3\,\s^{-1}$, and a possible origin of
	the observed behavior is discussed.
\end{abstract}

\vspace{2pc}

\maketitle

\section{Introduction} 

Experiments with ultracold atomic gases in combination with optical lattices
are a cornerstone in the investigation of quantum matter with, among others,
applications in quantum simulation and many-body
physics~\cite{bloch_many-body_2008}. While these single-component quantum gas
systems are essentially defect-free, the investigation of multi-component
quantum gases allows for a quantum simulation of phenomena requiring
impurities~\cite{massignan_polarons_2014}. As such basic Anderson
localization~\cite{anderson_absence_1958} phenomena, Anderson's orthogonality
catastrophe~\cite{anderson_infrared_1967} or Kondo
physics~\cite{kondo_resistance_1964} might be addressed. In this context
interest sparked in quantum degenerate mixtures of bosonic ytterbium (Yb) and
fermionic lithium (Li) as a prime candidate to experimentally implement
impurity systems. In addition, when forming dimers built-up of Yb and Li due
to the combination of alkaline-earth-like atoms (Yb) and alkali ones (Li)
spin-doublet molecules~\cite{roy_photoassociative_2016}, building blocks of
envisioned spin-lattice quantum simulators~\cite{micheli_toolbox_2006}, can be
realized. More recently the exploration of mixed-dimensional
systems~\cite{nishida_phases_2010} and even topological
superfluids~\cite{caracanhas_fermi-bose_2017} have also been proposed.

Common to all applications of an ultracold Yb-Li mixture system is a mandatory
good understanding of the interspecies interactions. Even 6 years after the
first successful demonstration of quantum degenerate mixtures of bosonic
ytterbium (Yb) and fermionic lithium (Li)~\cite{hansen_quantum_2011,
hara_quantum_2011} both a theoretical treatment of basic collisional
properties in the Yb-Li mixture system and their experimental determination
remain a challenging topic. In this respect important steps have been taken
since the first experimental realizations. While initially only the absolute
value of the interspecies ground state scattering length could be
determined~\cite{hara_quantum_2011, hansen_quantum_2011} recent efforts could
also confirm the interaction to be repulsive~\cite{roy_two-element_2017}.
After it was shown in a first theoretical treatment that the Yb-Li ground
state system probably does not support broad enough Feshbach
resonances~\cite{chin_feshbach_2010} that are easily exploited
experimentally~\cite{brue_magnetically_2012} a later work provided hints at
possibly usable Feshbach resonances involving the metastable $\triplet$ state
of Yb and the ground state of Li~\cite{gonzalez-martinez_magnetically_2013}.
It also recognized the importance of the anisotropy induced spin dependence in
the involved interspecies interactions. Those calculations have been further
pursued in later works~\cite{petrov_magnetic_2015, chen_anisotropy_2015} and
first experimental results~\cite{dowd_magnetic_2015, konishi_collisional_2016}
followed. 

In the research detailed in the present paper we study inelastic collisional
properties between localized ${}^{174}$Yb($\triplet$) atoms immersed in a
Fermi sea of $^6$Li. We employ a species specific, three-dimensional (3D)
optical lattice and control the internal states of Yb by using selective
excitation and those of Li by means of optical pumping. This allows for the
first time a systematic study of the spin dependence in collisions between
two-electron atoms in the metastable $\triplet$ state and alkali atoms in the
ground state. In previous experiments~\cite{konishi_collisional_2016} we
studied the inelastic collisions between ${}^{174}$Yb($\triplet$, $m_J = \{-2,
0\}$) and ${}^6$Li in the $F=1/2$ ground state manifold. There, we provided
detailed information on the inelastic loss rate coefficients, excluded spin
changing collisions as dominant processes and generally found no significant
differences between collisions involving the $m_J = -2$ or the $m_J = 0$ state
of Yb. In the present work we investigate the full range of Yb($\triplet$)
Zeeman sublevels, $m_J = -2,\ldots, +2$, and four states of Li, Li($\Li$, $F =
1/2$, $m_F = \pm 1/2$) and Li($\Li$, $F = 3/2$, $m_F = \pm 3/2$). We find
nearly identical inelastic collision rates for all combinations involving
Li($F=1/2$) and strongly state dependent rates for collisions with Li($F=3/2$)
atoms that vary by more than an order of magnitude. The present work unveils
another piece of information on the nature of the ultracold Yb-Li collisional
system and offers further insights into anisotropy induced relaxation
processes in collisions involving non-S-state
atoms~\cite{reid_fine-structure_1969, mies_molecular_1973,
krems_electronic_2004, hancox_magnetic_2004}.


\section{Experimental procedure}

The experiment proceeds along the same lines as presented
in~\cite{konishi_collisional_2016}. Briefly, a mixture of
quantum degenerate ${}^{174}$Yb and ${}^6$Li is prepared by forced evaporative
cooling in a crossed optical far-off-resonance trap. Different to our previous
works we introduce during the initial phase of the evaporative cooling an
optical pumping step to prepare a spin-polarized Li sample in either the $F =
1/2, m_F = \pm 1/2$ or $F = 3/2, m_F = \pm 3/2$ manifold of the ground state.
In the former case a $0.5~\msec$ pulse of circularly polarized light resonant to
the Li $F=1/2 \rightarrow F'=1/2$ D1-line transition together with light
resonant to the Li $F=3/2 \rightarrow F'=5/2$ D2-line transition is applied.
In the latter case light on the D1($F=3/2 \rightarrow F'=3/2$)-line and
D2($F=1/2 \rightarrow F'=3/2$)-line is used. By a suitable choice of a
homogeneous magnetic bias field we can thus prepare a spin-polarized Li sample
in any of the four states given above. The purity of the sample is verified by
standard time-of-flight absorption imaging where the atoms expand for
$1.2~\msec$ in a strong magnetic field gradient and found to be above 90\%.
Care is taken to maintain a sufficiently strong bias field during the
remainder of the experimental sequence of about $7~\G$ not to lose the state
of polarization. We choose the parameters of the experiment such as to
typically obtain a Bose-Einstein condensate of $10 \times 10^4$ Yb atoms and a
Fermi degenerate gas of $3 \times 10^4$ spin-polarized Li atoms. The
temperature of Li is $T_{\rm Li} \approx 300~{\rm nK}$ and $T_{\rm Li}/T_{\rm
F} \approx 0.2$, where $T_{\rm F}$ is the Fermi temperature.

\begin{figure}[bt]
\centering
	\includegraphics[width=\textwidth]{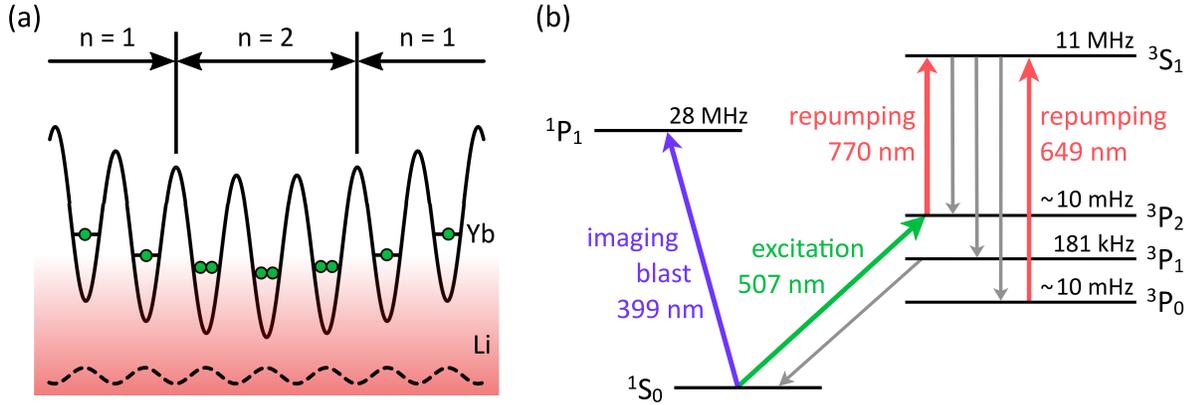}
\caption{
(a) Principle of the experimental method. In an optical lattice (solid line) operating at
wavelength $\lambda_{\rm L} = 532~\nm$ and depth $15\ E_R^{\rm Yb}$ the Yb
atoms (green dots) form a Mott insulator structure and are localized. For
clarity only lattice sites with occupation numbers $n = 1$ and $2$ are shown.
In the experiment the $507~\nm$ excitation light is tuned to only excite
singly occupied lattice sites to the desired $\triplet$ state. Li atoms (red)
only experience a small density modulation in the corresponding shallow optical
lattice (dashed line) at $0.7\ E_R^{\rm Li}$ and are delocalized.
(b) Basic level structure of Yb of relevance to the experiment. The
transitions for imaging and removal of ground state Yb atoms ($399~\nm$,
blue), for excitation to the $\triplet$ state ($507~\nm$, green) and repumping
to the ground state ($649~\nm$ and $770~\nm$, red) together with their
respective natural linewidths are indicated.
}
\label{fig:spectroscopy}
\end{figure}

We then proceed to adiabatically load the quantum degenerate mixture into a 3D
optical lattice with wavelength $\lambda_{\rm L} = 532~\nm$ and depth $15\
E_R^{\rm Yb}$, with $E_R^{\rm Yb}$ being the recoil energy of Yb in the
lattice, where Yb forms a Mott insulating state~\cite{bloch_many-body_2008,
konishi_collisional_2016}, see \fref{fig:spectroscopy}(a). In the same
configuration the corresponding lattice depth for Li is $0.7\ E_R^{\rm Li}$,
where $E_R^{\rm Li}$ is the Li recoil energy. The sign of the polarizability
of Li at $\lambda_{\rm L}$ is negative and the lattice sites of Yb and Li
alternate. As in our previous experiment spatial overlap between the two
atomic clouds is enhanced by use of a gravitational sag compensation beam that
is applied while the lattice is ramped up within $200~\msec$ to its target
power. The final separation between the Yb and Li cloud center-of-mass
positions is about $3.5~\micron$. Additionally, we adjust our magnetic bias
field to $200~\mG$ during the first $100~\msec$ of the lattice ramp and we
verified that the spin polarization of Li is maintained during this change.
The bias field lifts the degeneracy of the Yb($\triplet$) Zeeman states and a
resonant laser pulse of wavelength $507~\nm$ and duration $0.5~\msec$ excites
a small fraction of the Yb atoms to the desired $m_J$ Zeeman state. Usage of
the ultranarrow transition connecting the $\singlet$ to the metastable
$\triplet$ state, see \fref{fig:spectroscopy}(b), also allows us to
selectively only excite Yb atoms in singly occupied lattice
sites~\cite{konishi_collisional_2016}, thus suppressing inelastic decay from
collisions with other Yb($\triplet$) and Yb($\singlet$)
atoms~\cite{uetake_spin-dependent_2012}. During the $0.5~\msec$ excitation
time the excitation laser frequency is linearly ramped from $-4~{\rm kHz}$ to
$+4~\rm{kHz}$ around the resonance condition to ensure stable excitation even
with slight magnetic field noise due to background magnetic field changes in
the laboratory. The Zeeman splitting is $2.1~{\rm MHz}\,\G^{-1} \times h
\times m_J$, where $h$ is the Planck constant. The intensity of the excitation
light is chosen such as to excite about $2\times10^3$ to $3\times10^3$ atoms
to the $\triplet$ state, corresponding to about 10\% of the total number of Li
atoms. This ensures that the excited Yb atoms can be considered to be immersed
in a Fermi sea of Li atoms, i.e.\ the number of Li atoms can be considered as
constant during the remainder of the experiment. Remaining ground state Yb
atoms are removed within $0.3~\msec$ by application of light at $399~\nm$
resonant to the strong $\singlet \rightarrow {}^1\mathrm{P}_1$ transition.
This removal process ensures that also for magnetically sensitive states with
$m_J \neq 0$, where spurious excitation in lattice sites with higher
occupation numbers due to magnetic field fluctuations is possible, a clean
sample of strictly singly occupied Yb($\triplet$) lattice sites is prepared.
After a variable holding time an identical $399~\nm$ cleaning pulse is applied
and the remaining $\triplet$ atoms are repumped to the ground state where they
are recaptured by a magneto-optical trap operating also on the $\singlet
\rightarrow {}^1\mathrm{P}_1$ transition for fluorescence imaging detection
(see~\cite{konishi_collisional_2016} for details). The experimental signal is
thus the number of repumped Yb($\triplet$) atoms that remain in the optical
lattice after the holding time. By virtue of the second cleaning pulse we are
sensitive to all possible Yb($\triplet$) decay channels and measure the actual
number of metastable atoms remaining after the holding time. For Yb atoms in
the $\singlet$ state its depth is $15\ E_R^{\rm Yb} = 2.9~\mu{\rm K}\, k_{\rm
B}$ and is for $\triplet$ excited state atoms a factor $1$--$1.4$ deeper,
depending on the $m_J$ state.

\section{Results} 

\begin{figure}[bt]
\centering
\includegraphics[width=\textwidth]{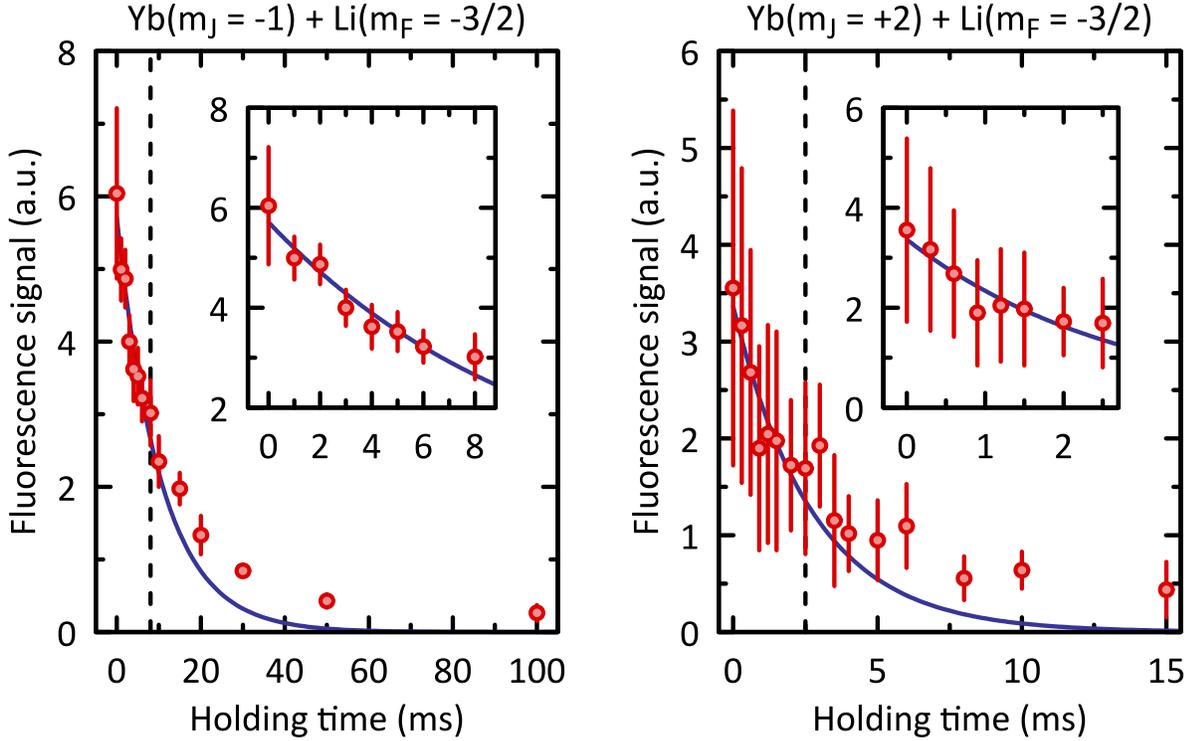}
\caption{Decay of Yb($\triplet$) atoms under inelastic collisions with Li. We
	record at $200~\mG$ bias field the decay of ${\rm Yb}(\triplet, m_J = -1)$
	atoms (left panel) and  ${\rm Yb}(\triplet, m_J = +2)$ atoms (right panel)
	under inelastic collision with ${\rm Li}(\Li, F = 3/2, m_F = -3/2)$ atoms.
	The intensity of the obtained fluorescence signal as a function of the
	holding time is shown. Note the different time ranges in both panels. The
	data points (red) are obtained from the mean of independent measurements
	at each holding time. The error bars represent the standard deviation
	thereof. An exponential decay function (blue, solid line) is fitted to the
	initial part of the data (indicated by a dashed line) and shown enlarged in
	the insets. The obtained lifetimes are $(10.4 \pm 1.3)~\msec$ and $(2.7 \pm
	0.9)~\msec$ for the left and right data set, respectively.}
\label{fig:decay}
\end{figure}

We systematically measure the inelastic Yb($\triplet$)-Li collisional
properties for all combinations of available $\triplet$ Zeeman states, $m_J =
-2, \ldots, +2$, and accessible Li ground states, $F = 1/2, m_F = \pm 1/2$ and
$F = 3/2, m_F = \pm 3/2$. For each combination of collisional partners we
record the decay of Yb($\triplet$) atoms by repeating the experiment several
times at various holding times. Typically about 15 different holding times
suitable for the observed speed of decay are chosen and at each holding time 5
(10) datapoints are taken for $|m_J| < 2$ ($|m_J| = 2$) states. More
datapoints are taken for measurements involving $|m_J| = 2$ states as due to
the high magnetic field sensitivity of those states, $4.2~{\rm MHz}\,\G^{-1}$,
data quality is reduced by inevitable magnetic field noise. Two typically
obtained decay curves are shown in \fref{fig:decay}. The recorded decay
behavior can be divided into two regimes. An initial relatively fast decay is
followed by notably slower dynamics. We attribute the qualitative change to a
transition from a Yb($\triplet$)-Li inelastic collisional dominated decay to a
mixed dynamics where both interspecies collisions and losses due to collisions
with thermal atoms and background gas are of importance.

The observed decay is attributed to be mostly caused by inelastic collisions
with Li atoms and to a minor extent due to collisions with background gas
atoms. Collisions between two Yb($\triplet$) atoms are suppressed as their
mobility is strongly reduced by the deep optical lattice. This motivates a
decay model~\cite{konishi_collisional_2016} for the Yb density, $n_{\rm Yb}$, 
\begin{equation}
	\dot{n}_{\rm Yb}({\bf r}, t) = - \alpha\, n_{\rm Yb}({\bf r}, t) - \beta\,
	\xi\, n_{\rm Li}({\bf r})\, n_{\rm Yb}({\bf r}, t)\, ,
	\label{eqn:DGL}
\end{equation}
where $\alpha$ is the one-body loss rate and $\beta$ the Yb($\triplet$)-Li
inelastic loss coefficient. The slight modulation of the Li density $n_{\rm Li}$
by the optical lattice is accounted for by the density correction factor
$\xi$. The correction factor is determined by the overlap of the Yb Wannier
state and the Li Bloch state at the respective lattice depths. While the lattice
depths depend on the Zeeman state dependent polarizabilities the overlap
integral only weakly changes~\cite{konishi_collisional_2016}. Throughout the
current work a constant value of $\xi = 0.65 \pm 0.03$ is adopted. Considering
the strong imbalance in the number of Li and Yb($\triplet$) atoms the density
$n_{\rm Li}({\bf r})$ is taken to be constant in time. The one-body loss rate
$\alpha$ cumulatively describes loss of Yb($\triplet$) atoms by spontaneous
decay and by inelastic collisions with background gas atoms. It is determined
by independent measurements in which the Li atoms have been removed from the
sample by a light pulse resonant to the Li D2 line. We typically observe
$\alpha^{-1} = (850 \pm 300)~\msec$. The complete decay is then described by
\begin{equation}
	N_{\rm Yb}(t) = \int n_{\rm Yb}({\bf r}, 0)\, e^{-(\alpha + \beta\xi
		n_{\rm Li}({\rm r}))\, t}\, {\rm d}^3r\, .
	\label{eqn:integral}
\end{equation}
Even though it was shown that the complete experimentally observed decay can be
described by \eqref{eqn:integral} we here adopt a different approach to the
analysis of the data. The initial loss of the total number of Yb($\triplet$)
atoms, $N_{\rm Yb}$, is accessible by spatial integration of \eqref{eqn:DGL},
\begin{equation}
	\dot{N}_{\rm Yb}(t = 0) = - \alpha\, N_{\rm Yb}(t=0) - \beta\, \xi\, \int
	n_{\rm Li}({\bf r})\, n_{\rm Yb}({\bf r}, 0)\, {\rm d}^3r\, .
	\label{eqn:initialDecay}
\end{equation}
For short holding times the loss of Yb($\triplet$) atoms is further
excellently described by an exponential decay behavior. Accordingly, 
we describe the initial decay of the data by
\begin{equation}
	N_{\rm Yb}^{\rm expt}(t) = N_{\rm Yb}(0)\, e^{-t/\tau_{\rm expt}}\, .
	\label{eqn:Nexpt}
\end{equation}
Comparison of this expression to \eqref{eqn:initialDecay} at $t = 0$
then gives access to the inelastic loss rate
\begin{equation}
	\beta = \frac{N_{\rm Yb}(0)}{\xi\, X} \left( \frac{1}{\tau_{\rm expt}} -
	\alpha \right)\, ,
	\label{eqn:beta}
\end{equation}
where we have introduced the overlap integral $X = \int
n_{\rm Li}({\bf r})\, n_{\rm Yb}({\bf r}, 0)\, {\rm d}^3r$. This allows for a
precise, stable and numerically fast analysis of data even under the
influence of experimental noise, as illustrated in \fref{fig:decay}.

Special care is taken for a sound treatment of the statistical and systematic
errors in the data analysis. This necessity stems in particular from the
pronounced sensitivity of the Yb($\triplet, m_J = \pm 2$) states to magnetic
noise. First, in a bootstrap approach the initial $5$--$10~\msec$ of available
data of each set is randomly resampled. That is, for each holding time a
number of points is randomly drawn from the set of experimental data so that a
new realization is obtained with the same number of data points as in the
original set at each time step, that is a single point is allowed to be drawn
more than once. This realization is then fitted by \eqref{eqn:Nexpt} and the
complete resampling sequence is repeated 2000 times. Thus determined
probability distribution function (PDF) of lifetimes $\tau_{\rm expt}$ then
serves as input to a second step in which \eqref{eqn:beta} is solved, again
2000 times, where in each case new representative values for each parameter
are randomly drawn from a given PDF. The obtained PDF of inelastic loss
coefficients is then expressed in terms of an cumulative distribution function
where the quantile at 50.0\%, i.e.\ the median, is taken as best estimate and
the quantiles at 15.9\% and 84.1\% serve as bounds for a 68.3\%, i.e.\ a
1-$\sigma$, confidence interval. The assumed statistical and systematic errors
are listed in table~\ref{tbl:errors}.

\begin{table}[bt]
	\caption{Statistical and systematic errors accounted for in the data
		analysis. The relative vertical distance between the Yb and Li atom cloud
		due to gravitational sag is denoted by $\delta_z$, possible relative cloud
		position offsets in horizontal direction are $\delta_{\{x, y\}}$. The
		three offsets contribute to the overlap integral $X$.}
	\label{tbl:errors}
	\centering
\begin{tabular}{lll}
	\hline
	\hline
	Parameter & Error type & Assumed distribution \\
	\hline
	$\tau$          & Statistic & Determined by fits to experimental data\\
	$\alpha^{-1}$   & Statistic & Normal distribution, $\sigma$ from fit to data \\
	$N_{\rm Yb}(0)$ & Statistic & Normal distribution, $\sigma = 0.3\, N_{\rm Yb}(0)$ \\
	$N_{\rm Li}$    & Statistic & Normal distribution, $\sigma = 0.3\, N_{\rm Li}$ \\
	$\xi$           & Systematic & Uniform distribution between $[0.62, 0.68]$ \\
	$\delta_z$      & Systematic & Uniform distribution between $[2.5, 4.5]~\micron$ \\
	$\delta_x$, $\delta_y$ & Systematic & Uniform distribution between $[-0.5, +0.5]~\micron$ \\
	\hline
	\hline
\end{tabular}
\end{table}

\begin{figure}[bt]
\centering
	\includegraphics[width=\textwidth]{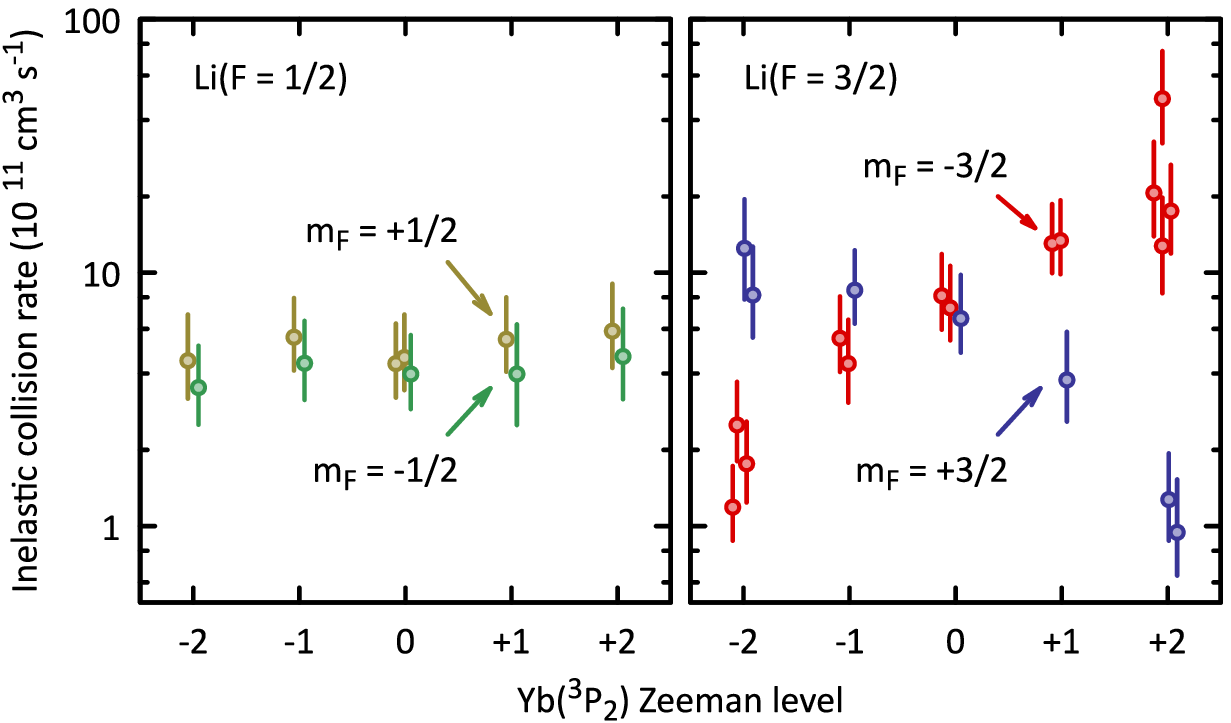}
\caption{Experimentally determined inelastic collision rates between
	Yb($\triplet$)-Li($\Li, F = 1/2$) (left panel) and Yb($\triplet$)- Li($\Li,
	F = 3/2$) (right panel) versus the $m_J$ Zeeman sublevel of Yb($\triplet$)
	at $200~\mG$ bias magnetic field. While inelastic collisions with Li($\Li, F
	= 1/2$) show no significant dependence on the $m_F$ magnetic state, in
	collisions with Li($\Li, F = 3/2$) a clear spin dependence is observed. For
	each measurement the median (points) and 1-$\sigma$ confidence interval
	(bars) is given (see main text for the details of the data analysis). To
	estimate data reproducibility various measurements have been repeated
	several times and we report all the data here. The data is slightly offset
	horizontally as needed for better visibility.}
\label{fig:collisionrates}
\end{figure}

The obtained inelastic collision rates are summarized in
\fref{fig:collisionrates}. Two distinct and different behaviors are observed:
In experiments with Li($F=1/2$) as collisional partner the inelastic collision
rate is constant at about $4\times10^{-11}~\cm^3\,\s^{-1}$. On the contrary,
in inelastic collisions with Li($F=3/2$) a strong, systematic dependence on
the choice of $m_J$ for Yb and $m_F$ for Li is found. Inelastic collision
rates vary between roughly $1\times10^{-11}$ and
$40\times10^{-11}~\cm^3\,\s^{-1}$. This is the first time that a dependence of
the collisional dynamics between metastable state two-electron atoms and the
ground state of alkali-atoms on the spin states of both species is confirmed
experimentally and poses the main result of this work. The symmetry of the
discovered spin dependence is striking. Nearly equally high inelastic
collision rates are found for Yb($\triplet, m_J = \mp 2$)-Li($F = 3/2, m_F =
\pm 3/2$) processes and nearly equally low rates are seen for Yb($\triplet,
m_J = \pm 2$)-Li($F = 3/2, m_F = \pm 3/2$) stretched state collisions. In the
intermediate regime, where $m_J = 0$, inelastic rates comparable to processes
including Li($F = 1/2$) are found.

\section{Discussion}

\begin{figure}[bt]
\centering
	\includegraphics[width=\textwidth]{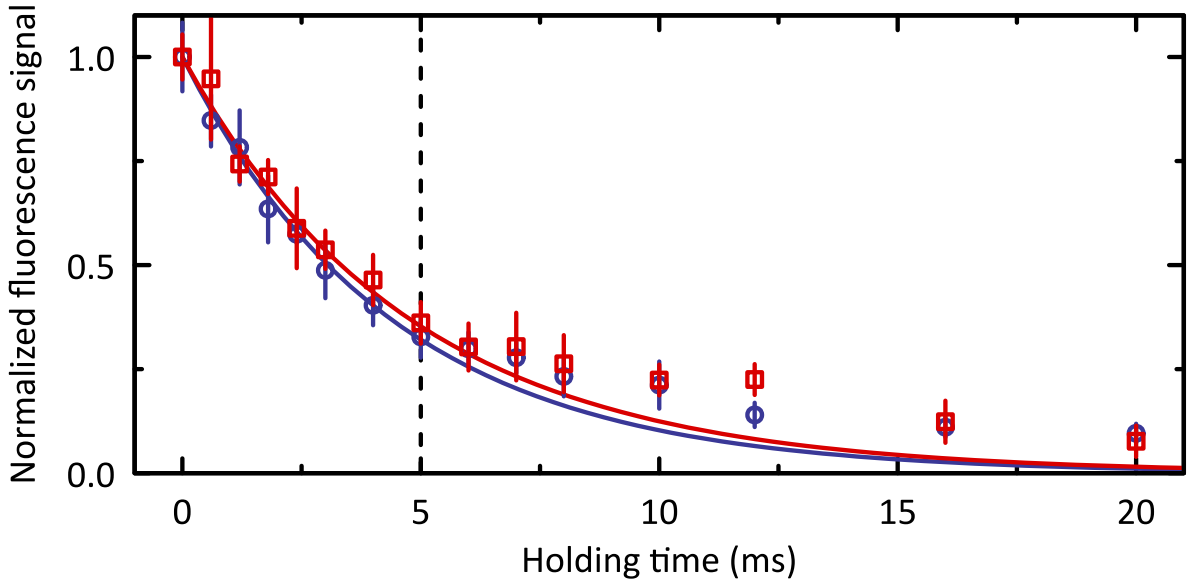}
\caption{Magnetic field dependence of Yb($\triplet, m_J = +1$)-Li($\Li, F =
3/2, m_F = -3/2$) collisional dynamics. Shown are the results obtained at
$200~\mG$ (blue circles) and at $100~\mG$ (red squares). The recorded
fluorescence signal is normalized to unity here for ease of comparison.
Exponential decay curves are fitted to the data up to including $5~\msec$
(dashed line). Observed lifetimes are $(4.4 \pm 0.2)~\msec$ and $(4.8 \pm
0.2)~\msec$ respectively and no significantly different decay behavior is
observed. This implies that Yb($\triplet$) spin changing collisions can be
excluded as dominant collisional process.}
\label{fig:spinflips}
\end{figure}

First, we want to shed some light on the possible inelastic decay channels of
importance for the collisional processes at hand. One distinguishes between
(i) spin changing, (ii) fine-structure changing, (iii) hyperfine-structure
changing and (iv) principal quantum number changing collisions. Spin changing
collisions are to be understood as processes where $m_J$ or $m_F$ change,
fine-structure changes imply a decay Yb($\triplet$) $\rightarrow$ Yb($^3{\rm
P}_1$ or $^3{\rm P}_0$), hyperfine structure changes account for Li($F = 3/2
\rightarrow 1/2$) processes and principal quantum number changes indicate a
direct Yb($\triplet$) $\rightarrow$ Yb($\singlet$) decay. In inelastic
collisions between Yb and Li about $m_{\rm Li}/(m_{\rm Yb} + m_{\rm Li})
\approx 3\%$, where $m_{\rm Yb, Li}$ denotes masses, of the released energy is
transferred onto Yb. As stated before our optical lattice has for Yb a depth
of at least $2.9~\mu{\rm K}\, k_{\rm B}$. This is to be compared to the energy
gain of Yb($\triplet$) atoms in a collisional process with $m_J \mapsto m_J -
1$ which is $0.6~\mu{\rm K}\, k_{\rm B}$ at $200~\mG$ considering the Yb-Li
kinematic factor $0.03$. Thus at least in $m_J = +2 \mapsto -2$ processes an
energy gain of about $2.4~\mu{\rm K}\, k_{\rm B}$ might lead to the onset of
increased particle loss from the optical lattice. This could partially explain
the enhanced inelastic loss rate observed in Yb($\triplet, m_J = +2$)-Li($\Li,
F = 3/2, m_F = -3/2$) collisions. To exclude such a possibility in this
situation the experiment is repeated at a reduced bias field of $100~\mG$ and
a comparison of both decay dynamics is shown in \fref{fig:spinflips}. No
significant differences that might hint at different loss mechanisms are
observed. This is in line with earlier reports of negligible spin flip
processes at low magnetic fields in collisions with Li($\Li, F = 1/2$)
atoms~\cite{dowd_magnetic_2015, konishi_collisional_2016}. Also, the
observation of strong inelastic losses in Yb($\triplet, m_J = -2$)-Li($F=3/2,
m_F = 3/2$) collisions, where exothermic spin changes of Yb are not possible,
leads to the same conclusion. Note that for Li a change $m_F \mapsto m_F-1$
only heats up Yb by $0.27~\mu{\rm K}$ and does not cause trap loss at both
magnetic fields considered. However, a hyperfine-structure changing event will
heat Yb by $330~\mu{\rm K}$ and surely lead to loss of Yb.

We now focus on the magnetic sublevel dependence of the observed inelastic
collision rates (see \fref{fig:collisionrates}). In short, while constant
relaxation rates are seen with Li($\Li, F = 1/2$) as collisional partner, for
Li($\Li, F = 3/2$) suppression of Zeeman
relaxation~\cite{krems_electronic_2004, hancox_magnetic_2004} in the stretched
state system and systematically increased inelastic collision rates for other
Yb($\triplet$) Zeeman sublevels are found. While a complete description of the
system Hamiltonian can be found for example
in~\cite{gonzalez-martinez_magnetically_2013} for the following discussion it
will be sufficient to only consider $\hat{U}(R)$, the interspecies interaction
operator as function of the relative distance $R$ between the Yb($\triplet$)
and Li atoms. The interaction produces four molecular states ($^2\Sigma^+$,
$^2\Pi$, $^4\Sigma^+$ and $^4\Pi$) that in the limit $R \rightarrow \infty$
dissociate to the Yb($^3{\rm P}$)-Li($\Li$) states. The four potential curves
are shown in~\cite{gonzalez-martinez_magnetically_2013, petrov_magnetic_2015}
to differ significantly at intermediate distances and, accordingly,
considerable Zeeman relaxation processes are
expected~\cite{krems_suppression_2005}. Therefore, the interplay between
anisotropic interaction induced decay and total spin conserving processes,
where the latter is the usual condition for alkali dimers, should lead to spin
dependent and enhanced relaxation mechanism. Indeed, our inelastic collision
rates are three to four orders of magnitude larger than those reported for,
e.g., Ti($^3{\rm F}_2$)-He collisions~\cite{hancox_suppression_2005} where
anisotropy induced effects are suppressed due to screening by outer $4 s$
orbitals. Lack of such a screening mechanism in the Yb($\triplet$)-Li system
is generally confirmed by our data. 

The interspecies interaction conserves $M_{\rm tot} = m_{J_{\rm Yb}} +
m_{F_{\rm Li}} + m_l$, the sum of the Yb and Li angular momentum projections
onto the axis of the applied magnetic field and $m_l$, the projection of the
collisional channel angular
momentum~\cite{gonzalez-martinez_magnetically_2013, petrov_magnetic_2015}.
Considering the case $m_l = 0$ conservation of $M_{\rm tot}$ leads to a lack
of inelastic decay channels for collisions in a stretched state configuration.
Even though a complete absence of inelastic decay is not observed in our
experimental data a strong suppression is revealed. More precisely, the
observed losses are due to the anisotropy in $\hat{U}(R)$.
In~\cite{krems_electronic_2004} the concept of internal and external
anisotropy in collisions with non-S-state atoms was introduced as leading and
higher order terms, respectively, in a tensorial expansion of the
Born-Oppenheimer potential. The internal anisotropy part of $\hat{U}$ does not
couple electronic angular momentum to the rotational angular momentum of the
nuclei and drives transitions 
$
| j_{\rm Yb}, m_{j_{\rm Yb}} \rangle \, 
| f_{\rm Li}, m_{f_{\rm Li}} \rangle \to 
| j'_{\rm Yb}, m_{j_{\rm Yb}} + \Delta m \rangle \, 
| f'_{\rm Li}, m_{f_{\rm Li}} - \Delta m \rangle
$.
Here, $j_{\rm Yb}$ and $f_{\rm Li}$ denote the atomic angular momenta of the
respective atomic species, their projections being expressed as $m_{j_{\rm
Yb}}$ and $m_{f_{\rm Li}}$. Internal anisotropy preserves total electronic
angular momentum and its projection. This implies that internal anisotropy
cannot drive transitions in stretched state collisions. In contrast to this,
the external anisotropy part couples to the rotational momentum of the nuclei
and causes transitions 
$
| j_{\rm Yb}, m_{j_{\rm Yb}} \rangle \, 
| f_{\rm Li}, m_{f_{\rm Li}} \rangle \to
| j'_{\rm Yb}, m_{j'_{\rm Yb}} \rangle \, 
| f'_{\rm Li}, m_{f'_{\rm Li}} \rangle
$.
It does not preserve electronic angular momentum and therefore
allows for transitions in stretched state collisions. Our stretched state
results ($M_{\rm tot} = \pm 7/2$) estimate in very good agreement with the
theoretical result~\cite{gonzalez-martinez_magnetically_2013} the external
anisotropy induced inelastic collision rate to about $1 \times
10^{-11}~\cm^3\,\s^{-1}$. In contrast, in non-stretched state collisions not
only external anisotropy but also internal anisotropy leads to relaxation
effects. The strong impact of the internal anisotropy is underlined by the
more than tenfold enhanced inelastic collision rates observed for $M_{\rm tot}
= \pm 1/2$ that approach the predicted universal loss
rate~\cite{idziaszek_universal_2010, gonzalez-martinez_magnetically_2013} at
$2.9 \times 10^{-10}~\cm^3\,\s^{-1}$ indicative of complete loss at short range. As
demonstrated in~\cite{frye_cold_2015}, strong deviations from the universal
model can lead to pronounced fluctuations of the collisional cross-sections.

Turning our attention to inelastic processes involving Li($F = 1/2$) states
(see \fref{fig:collisionrates}, left) a structureless behavior is found. In
all spin combinations the inelastic collision rate is about $4 \times
10^{-11}~\cm^3\,\s^{-1}$, roughly equal to the mean behavior observed with
Li($F = 3/2$) for $1/2 < |M_{\rm tot}| < 7/2$. The reason for this stark
difference is not clear and no systematic theoretical investigation of this
particular case has yet been reported. In particular the role of possible
Li($F = 3/2$) $\rightarrow$ Li($F = 1/2$) relaxation processes is unknown. Due
to the large energy gain in such a hyperfine-level changing collision the
involved Li atom is surely lost from our trap and cannot be distinguished from
a Li $m_F$ changing event in the current experiment.


\section{Conclusions and outlook}

Investigating state selectively excited Yb($\triplet$) in a deep optical
lattice embedded in a sea of spin polarized Li we reported the interspecies
inelastic collision rates for different spin configurations of the
constituents at low magnetic fields. Of particular interest was the systematic
dependence of the Yb($\triplet$)-Li($F = 3/2$) inelastic collision rate on
$|M_{\rm tot}|$ where suppressed relaxation for the stretched states, $|M_{\rm
tot}| = 7/2$, and more than tenfold increased loss rates were found for
$|M_{\rm tot}| = 1/2$. While we could demonstrate those results to be
consistent with the ongoing research on collisions involving non-S-state atoms
a detailed understanding, in particular of the Yb($\triplet$)-Li($F = 1/2$)
collisional process, is as of yet missing. The presented data should thus
stimulate new theoretical efforts for a better grasp on the details of the
physical processes involved. At the same time the results indicate that in
particular the stretched state configurations might be noteworthy candidates
for future experiments dedicated to find means to tune the Yb($\triplet$)-Li
interspecies interactions by a Feshbach resonance effect. It will now be a new
challenge to take the presented results as a starting point for systematic
investigations at stronger magnetic bias fields so that finally a good
understanding of the Yb($\triplet$)-Li system might be achieved.


\ack
We acknowledge useful comments by J.\ M.\ Doyle, E.\ Chae and M.\ Kumakura.
This work was supported by the Grant-in-Aid for Scientific Research of JSPS
No.\ JP25220711, No.\ JP26247064, No.\ JP16H00990, and No.\ JP16H01053 and the
Impulsing Paradigm Change through Disruptive Technologies (ImPACT) program by
the Cabinet Office, Government of Japan. H.\ K.\ acknowledges support from
JSPS.

\section*{References}
\providecommand{\newblock}{}


\begin{thebibliography}{10}
\expandafter\ifx\csname url\endcsname\relax
  \def\url#1{{\tt #1}}\fi
\expandafter\ifx\csname urlprefix\endcsname\relax\def\urlprefix{URL }\fi
\providecommand{\eprint}[2][]{\url{#2}}

\bibitem{bloch_many-body_2008}
Bloch I, Dalibard J and Zwerger W  {\bf 80} 885--964 ISSN 0034-6861
  \urlprefix\url{http://link.aps.org/doi/10.1103/RevModPhys.80.885}

\bibitem{massignan_polarons_2014}
Massignan P, Zaccanti M and Bruun G~M  {\bf 77} 034401 ISSN 0034-4885
  \urlprefix\url{http://stacks.iop.org/0034-4885/77/i=3/a=034401}

\bibitem{anderson_absence_1958}
Anderson P~W  {\bf 109} 1492--1505

\bibitem{anderson_infrared_1967}
Anderson P~W  {\bf 18} 1049--1051

\bibitem{kondo_resistance_1964}
Kondo J  {\bf 32} 37--49 ISSN 0033-068X, 1347-4081
  \urlprefix\url{https://academic.oup.com/ptp/article-lookup/doi/10.1143/PTP.32.37}

\bibitem{roy_photoassociative_2016}
Roy R, Shrestha R, Green A, Gupta S, Li M, Kotochigova S, Petrov A and Yuen C~H
   {\bf 94} 033413
  \urlprefix\url{http://link.aps.org/doi/10.1103/PhysRevA.94.033413}

\bibitem{micheli_toolbox_2006}
Micheli A, Brennen G~K and Zoller P  {\bf 2} 341--347 ISSN 1745-2473
  \urlprefix\url{http://www.nature.com/nphys/journal/v2/n5/full/nphys287.html}

\bibitem{nishida_phases_2010}
Nishida Y  {\bf 82} 011605
  \urlprefix\url{http://link.aps.org/doi/10.1103/PhysRevA.82.011605}

\bibitem{caracanhas_fermi-bose_2017}
Caracanhas M~A, Schreck F and Smith C~M  (\textit{Preprint}
  \eprint{1701.04702}) \urlprefix\url{http://arxiv.org/abs/1701.04702}

\bibitem{hansen_quantum_2011}
Hansen A~H, Khramov A, Dowd W~H, Jamison A~O, Ivanov V~V and Gupta S  {\bf 84}
  011606 \urlprefix\url{http://link.aps.org/doi/10.1103/PhysRevA.84.011606}

\bibitem{hara_quantum_2011}
Hara H, Takasu Y, Yamaoka Y, Doyle J~M and Takahashi Y  {\bf 106} 205304
  \urlprefix\url{http://link.aps.org/doi/10.1103/PhysRevLett.106.205304}

\bibitem{roy_two-element_2017}
Roy R, Green A, Bowler R and Gupta S  {\bf 118} 055301
  \urlprefix\url{http://journals.aps.org/prl/abstract/10.1103/PhysRevLett.118.055301}

\bibitem{chin_feshbach_2010}
Chin C, Grimm R, Julienne P and Tiesinga E  {\bf 82} 1225
  \urlprefix\url{http://link.aps.org/doi/10.1103/RevModPhys.82.1225}

\bibitem{brue_magnetically_2012}
Brue D~A and Hutson J~M  {\bf 108} 043201
  \urlprefix\url{http://link.aps.org/doi/10.1103/PhysRevLett.108.043201}

\bibitem{gonzalez-martinez_magnetically_2013}
Gonz\'{a}lez-Mart\'{i}nez M~L and Hutson J~M  {\bf 88} 020701
  \urlprefix\url{http://link.aps.org/doi/10.1103/PhysRevA.88.020701}

\bibitem{petrov_magnetic_2015}
Petrov A, Makrides C and Kotochigova S  {\bf 17} 045010 ISSN 1367-2630
  \urlprefix\url{http://stacks.iop.org/1367-2630/17/i=4/a=045010}

\bibitem{chen_anisotropy_2015}
Chen T, Zhang C, Li X, Qian J and Wang Y  {\bf 17} 103036 ISSN 1367-2630
  \urlprefix\url{http://stacks.iop.org/1367-2630/17/i=10/a=103036}

\bibitem{dowd_magnetic_2015}
Dowd W, Roy R~J, Shrestha R~K, Petrov A, Makrides C, Kotochigova S and Gupta S
  {\bf 17} 055007 ISSN 1367-2630
  \urlprefix\url{http://iopscience.iop.org/article/10.1088/1367-2630/17/5/055007/meta}

\bibitem{konishi_collisional_2016}
Konishi H, Sch\"{a}fer F, Ueda S and Takahashi Y  {\bf 18} 103009 ISSN
  1367-2630 \urlprefix\url{http://stacks.iop.org/1367-2630/18/i=10/a=103009}

\bibitem{reid_fine-structure_1969}
Reid R~H~G and Dalgarno A  {\bf 22} 1029--1030

\bibitem{mies_molecular_1973}
Mies F~H  {\bf 7} 942--957

\bibitem{krems_electronic_2004}
Krems R~V, Groenenboom G~C and Dalgarno A  {\bf 108} 8941--8948 ISSN 1089-5639
  \urlprefix\url{http://dx.doi.org/10.1021/jp0488416}

\bibitem{hancox_magnetic_2004}
Hancox C~I, Doret S~C, Hummon M~T, Luo L and Doyle J~M  {\bf 431} 281--284 ISSN
  0028-0836
  \urlprefix\url{http://www.nature.com/nature/journal/v431/n7006/abs/nature02938.html}

\bibitem{uetake_spin-dependent_2012}
Uetake S, Murakami R, Doyle J~M and Takahashi Y  {\bf 86} 032712
  \urlprefix\url{http://link.aps.org/doi/10.1103/PhysRevA.86.032712}

\bibitem{krems_suppression_2005}
Krems R~V, Kłos J, Rode M~F, Szczȩśniak M~M, Chałasiński G and Dalgarno A
  {\bf 94} 013202 ISSN 0031-9007, 1079-7114
  \urlprefix\url{https://link.aps.org/doi/10.1103/PhysRevLett.94.013202}

\bibitem{hancox_suppression_2005}
Hancox C~I, Doret S~C, Hummon M~T, Krems R~V and Doyle J~M  {\bf 94} 013201
  ISSN 0031-9007, 1079-7114
  \urlprefix\url{https://link.aps.org/doi/10.1103/PhysRevLett.94.013201}

\bibitem{idziaszek_universal_2010}
Idziaszek Z and Julienne P~S  {\bf 104} 113202
  \urlprefix\url{http://link.aps.org/doi/10.1103/PhysRevLett.104.113202}

\bibitem{frye_cold_2015}
Frye M~D, Julienne P~S and Hutson J~M  {\bf 17} 045019 ISSN 1367-2630
  \urlprefix\url{http://stacks.iop.org/1367-2630/17/i=4/a=045019}

\end{thebibliography}
\end{document}